\documentstyle[12pt,aaspp4,epsf]{article}
\begin{document}

\title{Simulations of binary coalescence of a neutron star \\ and a black hole}

\author{W. Klu\'zniak and W.H. Lee}
\affil{University of Wisconsin--Madison, Physics Deparment, \\
1150 University Ave., Madison, WI 53706 \\
Copernicus Astronomical Center, ul. Bartycka 18, 00-716 Warszawa, Poland}

\lefthead{Klu\'{z}niak \& Lee}
\righthead{Black Hole--Neutron Star coalescence}

\begin{abstract}
We present the results of Newtonian hydrodynamic simulations of the
coalescence of a binary consisting of a black hole with a neutron
star. The calculations show that for a wide range of initial
conditions the core of the neutron star survives the initial mass
transfer episode. We therefore identify black hole--neutron star
binaries as the astrophysical production site of low mass neutron
stars unstable to explosion. The relevance of the simulations to the
theory of gamma--ray bursts is also discussed.
\end{abstract}

Subject headings: gamma rays: bursts --- binaries: close --- stars:
neutron --- hydrodynamics

\section{Introduction}

The ultimate fate of a close binary composed of a neutron star and a
black hole has first been discussed (\cite{wheel}) shortly after the
discovery of neutron stars. It has been pointed out that the
coalescence of such a binary would make a promising site for the
r--process nucleosynthesis (\cite{latt}); the same authors already
suggested that the coalescence may give rise to a gamma--ray burst
(GRB), but the correctly estimated event rate was thought to be too
low in the then prevailing paradigm of Galactic sources for GRBs. As
discussed in the next section, recent observations led to a revived
interest in black hole--neutron star binaries as sources of GRBs.

A seemingly separate problem is that of the fate of a neutron star
with mass below the stability limit
(e.g. \cite{page},~\cite{sumiyoshi}). It has been though that such stars
will undergo a violent explosion but no reliable production sites had
been identified.

In this Letter we report on our Newtonian simulations of the final
stages of evolution of a black hole--neutron star
binary. Surprisingly, our results suggest a possible unification of
the disparate paths of investigation mentioned above.

\section{Black hole--neutron star coalescence as a potential source of GRBs}

The properties of dim optical transients
(\cite{vanP},~\cite{djorg},~\cite{metzger}) associated with gamma-ray
bursts (GRBs) reinforce the view (\cite{bp},~\cite{meeg}), hitherto
held on statistical grounds, that the sources of the observed GRBs are
not located in the Galaxy or the nearby clusters of galaxies.  All
facts are consistent with a ``cosmological'' origin of GRBs
(\cite{fish}).  In fact, the isotropy of GRBs and the distribution of
their peak flux favour a typical distance between $\sim100\,$Mpc and
$\sim1\,$Gpc to the closest sources of the observed GRBs. The reported
redshift (\cite{metzger}) of $z=0.8$ to the optical counterpart of
GRB970508 should settle the issue of the intrinsic luminosity of the
GRB sources.  A distance of $\sim1\,$Gpc implies that up to $10^{51}$
ergs must be released in gamma rays to account for the observed
fluences of $\sim10^{-7.5}$ to $\sim10^{-3}$erg/cm$^2$. All models
(\cite{colg},~\cite{bpap},
~\cite{eich},~\cite{bp},~\cite{usov},~\cite{mesz},~\cite{woos})
involve the birth or death of a neutron star or a star like it.

To be efficiently converted to observed gamma rays, the energy
released in the primary event must have a line of sight to the
observer which is sufficiently baryon-free to allow a relativistic
blast wave (\cite{bpap},~\cite{mesz},~\cite{rees}) to expand at
velocities close to the speed of light. It has been argued
(\cite{rys}) that the interaction of such relativistic outflow with
the interstellar medium will result in shock acceleration of electrons
and amplification of magnetic fields yielding significant emission of
gamma-rays rays through synchrotron radiation. The expected afterglow
(\cite{viet}) may be consistent with the X-ray and optical transients
detected by the Beppo-SAX satellite and follow-up observations
(\cite{vanP}).  We are looking, then, for a process which would
release a sufficient amount of energy in a baryon-free direction, and
one whose characteristic timescales correspond to the variability and
durations of the observed GRBs (in the shocked fireball model the GRB
timescales must arise at the source (\cite{piran})).

A sufficiently small baryon loading of the plasma is obtained
(\cite{pa}) in a natural way in the mergers of two strange stars,
because the strange-quark matter making up their bulk is self-bound
and hence immune to lofting by radiation. But the disruption of a
strange star would pollute the Galactic environment with strange-quark
nuggets which would preclude (\cite{cald}) the further formation of
young pulsars (neutron stars). Thus, the merger of a strange star with
anything else is excluded as a source of GRBs (\cite{wk}).

The most conservative scenario of GRB formation involves the
coalescence of a binary system composed of two neutron stars
(\cite{bpap},~\cite{eich}). These events are certain to occcur and a
satisfactory lower limit to their rate can reliably be inferred
(\cite{latt},~\cite{nara}), e.g. from the statistics of the known
Hulse-Taylor type neutron star binaries. There is disagreement as to
the outcome of the last stages of evolution of such
binaries. Newtonian simulations give an insufficient neutrino
luminosity to power a GRB (\cite{ruff}) while general relativistic
calculations indicate no blast wave will be formed, although a GRB
with a smooth time profile is the computed outcome
(\cite{wilson},~\cite{wilson97}).

It has been proposed (\cite{bp}) that in the binary coalescence of a
neutron star with a black hole the star would be disrupted into a
torus which would accrete on the viscous timescale, thus extending the
duration of the burst. Our simulations show a rather different
outcome, but it remains true that the process is extended in time (for
a different reason). Theoretical estimates (\cite{latt},~\cite{nara})
give $\sim10^{-6}$ per year per galaxy for the rate of coalescence of
such binaries, in agreement with the observed rate of GRBs. The energy
release is comparable to that in the double neutron star
mergers. Thus, the process seems to share all the advantages of the
coalescing neutron stars scenario, while avoiding its main
shortcomings. This motivated our study.

\section{Numerical Method}

For the computations presented in this letter, we have used a fully
Newtonian smooth particle hydrodynamics (SPH) code
(\cite{Lucy},~\cite{GM}). A detailed description of the code will be
published elsewhere (\cite{longpaper}). In calibration runs of the
code, we have replicated (\cite{my}) in detail all features of the
binary neutron star mergers computed by \cite{rasio}. The neutron star
was modeled as a polytrope with a stiff equation of state (adiabatic
index $\Gamma=3$) with 17,000 particles. The black hole was modeled as
a point mass with an absorbing boundary at $r_{g}=2GM/c^{2}$. Any
particle that comes closer than $r_{g}$ to the black hole is absorbed,
the mass and momentum of the black hole are adjusted so the that total
mass and linear momentum are conserved. The detailed results presented
here were obtained for initial conditions corresponding to a tidally
locked neutron star. Initial synchronized equilibrium configurations
can be constructed via a relaxation technique for a range of binary
separations, allowing the polytrope to respond to the presence of the
tidal field (\cite{RS92}). During the dynamical coalescence, we also
calculate the gravitational radiation waveforms emmitted by the
system, in the quadrupole approximation. These waveforms are 
presented elsewhere (\cite{eamaldi},b).

\section{Results and Discussion}

In the coalescence, the two components of the binary are brought
together by the loss of angular momentum to gravitational radiation. A
particularly interesting case occurs when the mass of the black hole
is close to that of the neutron star. In this case a dynamical
instability appears, and the orbit decays on a dynamical
timescale. The results of model calculations with an initial mass of
the neutron star of 1.4 M$_\odot$ and unperturbed radius of the
polytrope of 13.4 km are presented in Figures \ref{fig1} and
\ref{fig2}. Upon relaxing the polytrope to a synchronized state in the
binary system, we find the onset of instability at a distance of 37
km, this is the initial binary separation in the simulation presented
in Figures \ref{fig1} and \ref{fig2}.  

Figure \ref{fig2} shows density contour snapshots during a dynamical
simulation with an initial mass ratio of one ($q=1$).  A transient
massive accretion torus forms around the black hole, but the neutron
star is not completely disrupted as a result of this encounter. To the
limit of our resolution ($10^{-4}$M$_{\odot}$), a baryon--free line of
sight, parallel to the rotation axis of the binary, remains present
throughout the simulation. Higher resolution runs are needed to
determine whether the baryon content is below $10^{-5}$M$_{\odot}$, as
required by the blast--wave model of GRBs (\cite{rees}). The total
energy released through viscous heating is $\approx 5\times 10^{52}$
erg. In this case, mass transfer is essentially over in approximately
five initial orbital periods (11 ms) and a remnant core containing
0.43 M$_{\odot}$ is left orbiting around a 2.25 M$_{\odot}$ black
hole.

In Figure \ref{fig1} we have plotted (solid line) the mass accretion
rate onto the black hole, showing that the accretion event is very
brief $\sim 2\,$ms; the dashed line is the mass of the black hole as a
function of time.  The configuration resulting from the unstable mass
transfer in a binary of initial mass ratio $q=1$ is that of a black
hole and a lighter remnant core left in orbit of greater separation
($\sim 60\,$km) and a greatly altered mass ratio ($q_{final}=0.19$).
The orbital separation in this new binary system will again decrease
due to continuing emission of gravitational waves and, after
$\sim0.1\,$s, Roche-lobe overflow will occur as described below. In
the initial mass transfer for $q=1$, the black hole was a messy eater
and $\sim 0.1$M$_{\odot}$ of mass from the original neutron star remains
scattered around the binary system.  With the current resolution of
our computations we were unable to determine the exact distribution of
this matter after 0.1 s.  It is possible that some of this matter will
eventually be accreted onto the black hole, potentially releasing up
to $10^{51}$ erg in energy. We expect that much of the remaining
neutron matter will release its nuclear binding energy on the beta
decay timescale ($\tau\approx 15$ minutes).

For mass ratios not too close to unity ($q\equiv M_{ns}/M_{BH}<0.8$),
we find no dynamical instability. Once the components are brought
sufficiently close an episode of mass transfer through Roche--lobe
overflow from the neutron star onto the black hole ensues. This causes
the neutron star to move away from the black hole (by conservation of
angular momentum). The event is ``clean,'' all the mass lost by the
neutron star is accreted by the black hole.  These results are quite
different from those of early estimates, which suggested that the
neutron star will be tidally disrupted (\cite{wheel},~\cite{latt}) and
that a few per cent of the mass will be ejected (\cite{latt}) to
infinity, although it should be noted that our calculations are
completely Newtonian. If gravitational radiation backreaction is
neglected, the peak accretion rate is about 2M$_\odot/$s, but only
about one percent of the neutron star mass is transferred in each
episode.

The accretion rate and the mass transferred in such an episode are
illustrated in Figure \ref{fig3}, for a 1.4M$_\odot$ neutron star
orbiting a 4.5M$_\odot$ black hole ($q=0.31$).  Here, the critical
distance corresponding to Roche--lobe overflow is 50.4 km. After an
interval of time comparable with the duration of the accretion event,
$\sim 4\,$ms, gravitational radiation again forces the binary into a
configuration where mass transfer occurs again. Clearly, the number of
such accretion events would be $\sim M_{ns}/\Delta M_{BH}\sim 100$ and
the total duration of the process a few seconds. However,
gravitational radiation losses cannot be ignored in this case, since
the time scale for decay for the orbit (from angular momentum losses
to gravitational waves) in the point mass approximation is 3.5 ms and
the duration of the mass transfer episode presented in Figure
\ref{fig3} is 10 ms. To explore how these angular momentum losses to
gravitational radiation will affect the binary, we have calculated,
using the quadrupole approximation for angular momentum loss, the
evolution of the same binary assuming that the gravitational potential
is that of two point masses. After 10 ms of mass transfer, the binary
separation has increased by about 0.06\% and the mass of the neutron
star is 0.85 M$_{\odot}$. Thus, this approximation also leads to the
conclusion that the binary will survive with an altered mass ratio and
separation, and the total time scale of the coalescence process is
extended from a few milliseconds to at least several tens of
milliseconds. Full hydrodynamical simulations involving a backreaction
force are required to explore the evolution of such a binary in
greater detail.

Note that we have identified the final stages of evolution of the
black-hole neutron-star binary as the only known astrophysical process
leading to the creation of a low-mass neutron star.  The coalescence
ends with an explosion (\cite{colp},~\cite{sumiyoshi}) when the mass
of the surviving core drops below the lower stability limit of neutron
stars. This in itself could also give rise to an observable
transient. As pointed out by the referee, the black hole member of the
binary will be left behind with a large linear velocity as a result of
the explosion and the associated recoil. Our simulations suggest a
velocity on the order of $10^{4}$ km/s.

In summary, we have identified several unexpected features in the
binary coalescence of a neutron star with a black hole, which may make
such events promising candidate sources for the central engine of
gamma-ray bursters, at least for the shorter bursts in the apparently
bimodal distribution (\cite{kouveliotou}). The Newtonian numerical
calculations presented here assumed that the rotation of the neutron
star was synchronized with the orbital period. In fact, tidal locking
is not expected (\cite{bild}). Our preliminary simulations for a
non--synchronized system with an initial mass ratio of $q=0.31$ show
that the core of the neutron star survives the initial mass transfer
episode and could be driven below the minimum mass required for
stability. Thus the outcome is similar to that for the tidally locked
binary.  Finally, all of our results are predicated on the assumption
that the neutron star will not collapse to a black hole before the
onset of mass transfer, relativistic simulations are required to
address the validity of this assumption (\cite{wilson}).

\acknowledgements

This work was supported in part by Poland's Committee for Scientific
Research under grant KBN 2P03D01311 and by DGAPA--UNAM. We thank the
referee for helpful comments.

\newpage

\begin{figure}

\caption{Black hole mass (dashed line) and mass accretion rate onto
the black hole (solid line) for an initial mass ratio of q=1. \label{fig1}}

\end{figure}

\begin{figure}

\caption{Density contours at various times during the dynamical
coalescence of the black hole--neutron star binary with an initial
mass ratio q=1 and initial separation r=37 km. The color--coded
density ranges from $5 \times 10^{17}$ kg m$^{-3}$ (bright yellow) to $5
\times 10^{14}$ kg m$^{-3}$ (dark blue), and the box shown is 161 km on
a side. The rotation is about the z--axis and counterclockwise in a)
and b); the initial orbital period is P=2.3 ms. Density contours in
the orbital plane are shown at: a) t=4.6 ms and b) t=6.9 ms. Contours
in the meridional plane are shown at: c) t=6.9 ms and d) t=9.2 ms. The
baryon--free axis and the transient accretion torus are clearly
seen. The black disk represents the black hole. \label{fig2}}

\end{figure}

\begin{figure}

\caption{The change in black hole mass (dashed line) and mass
accretion rate onto the black hole (solid line) for an initial mass
ratio of q=0.31. The turn--off of the mass transfer may be related to
the absence of gravitational radiation reaction in our
simulation. \label{fig3}}

\end{figure}

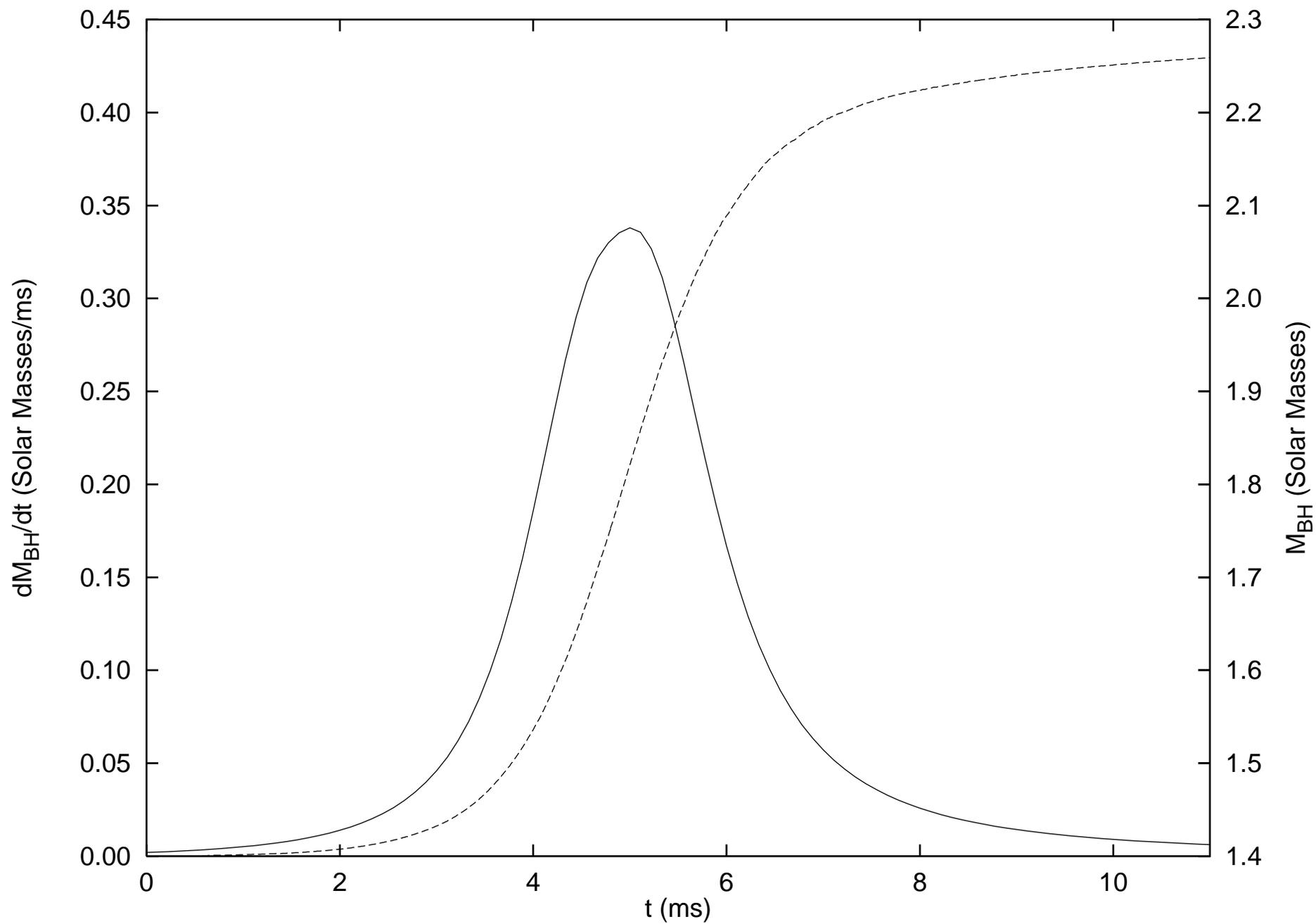
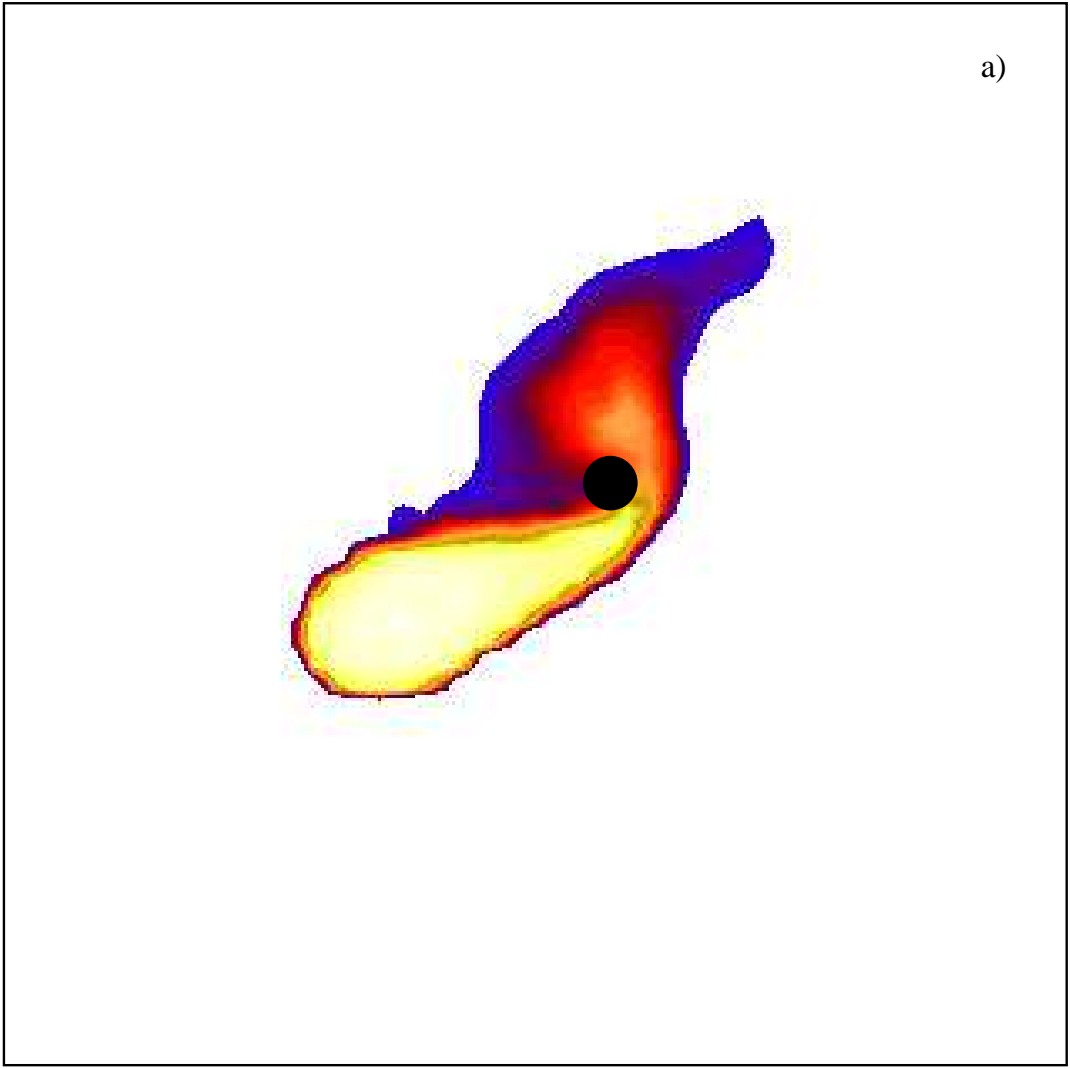
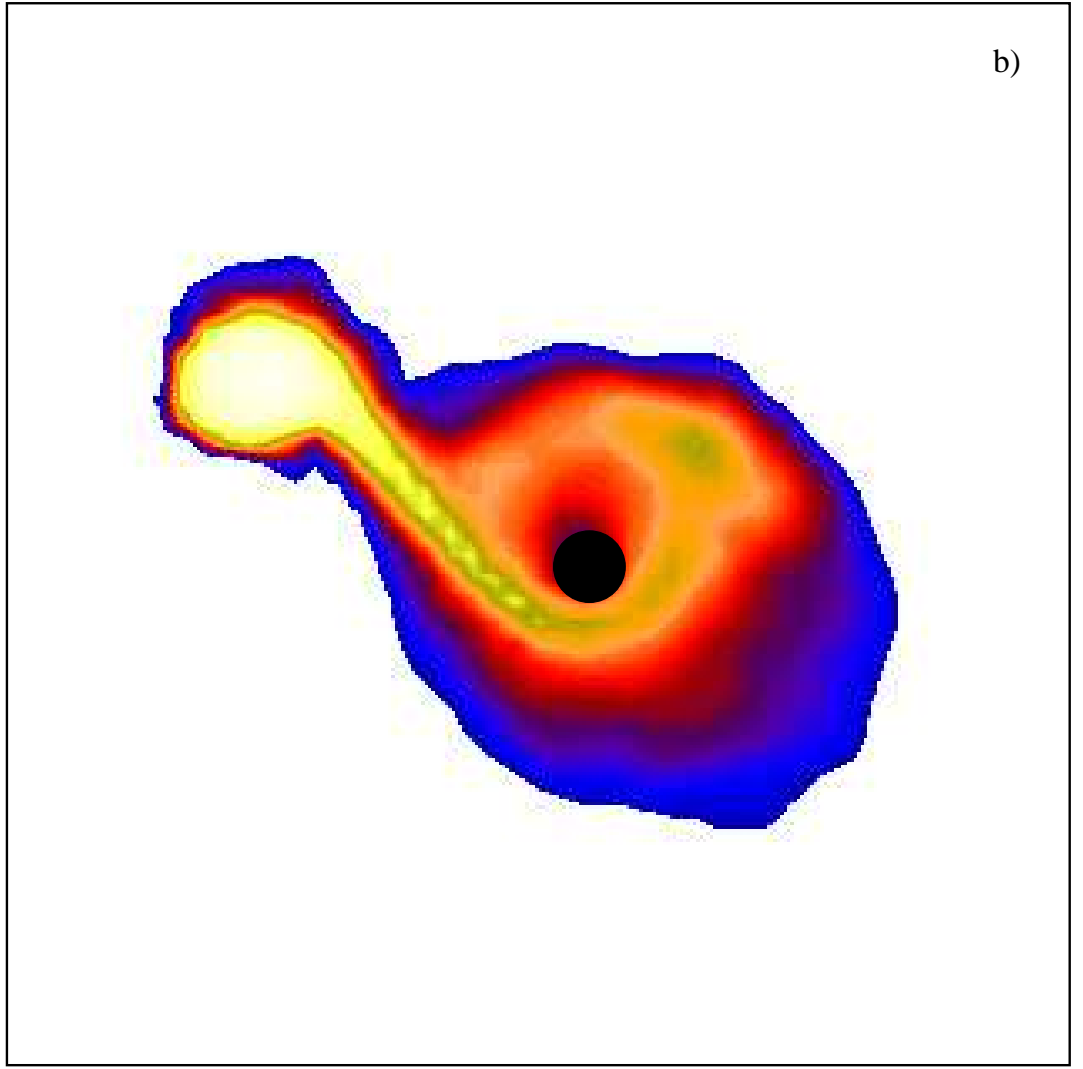
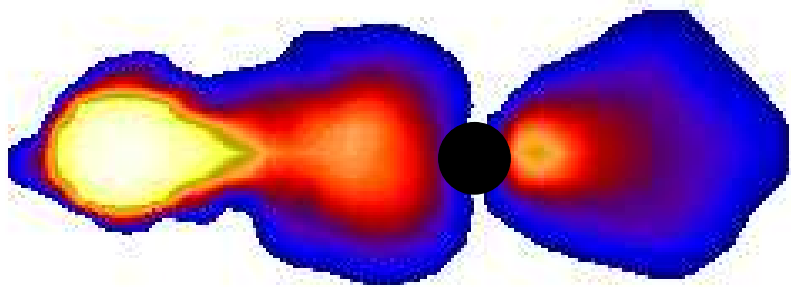
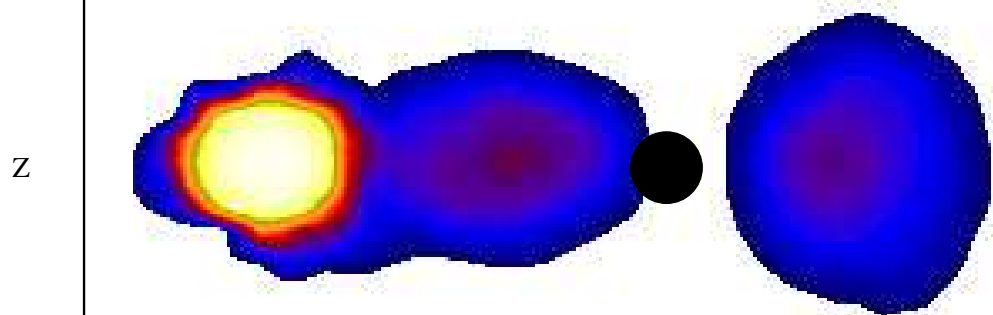
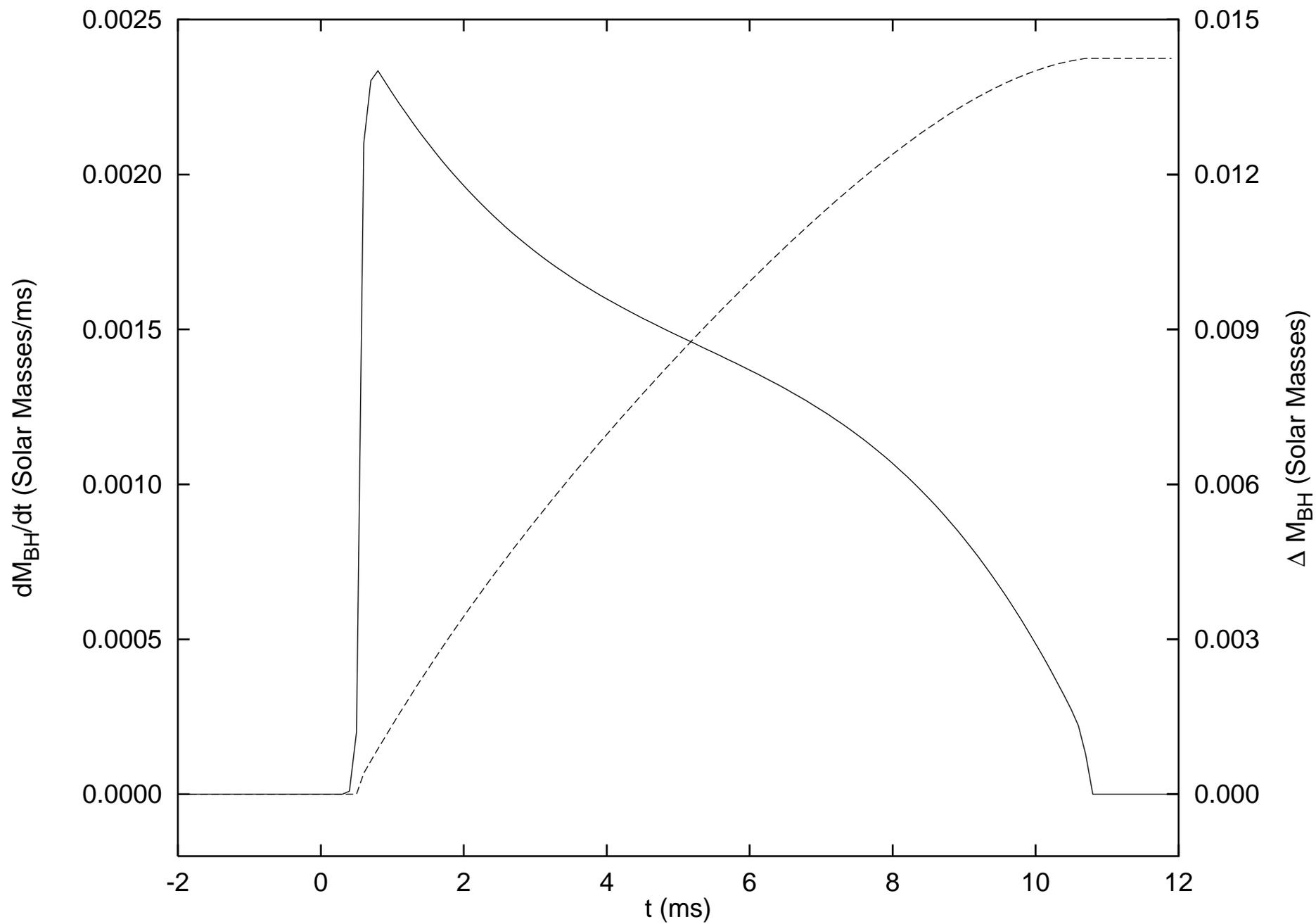
\end{document}